\documentclass[conference]{IEEEtran}
\IEEEoverridecommandlockouts
\usepackage{cite}
\usepackage{amsmath,amssymb,amsfonts}
\usepackage{algorithmic}
\usepackage{graphicx}
\usepackage{textcomp}
\usepackage{amssymb,amsfonts,mathtools}
\usepackage[lofdepth,lotdepth]{subfig}
\usepackage{bm}
\usepackage{balance,cite}
\usepackage{color, colortbl}
\usepackage{hyperref}
\usepackage{cleveref}
\usepackage{float}
\usepackage{algorithm}
\usepackage{wrapfig}
\usepackage{algorithmic}
\usepackage{multirow}
\usepackage{cuted}
\usepackage{verbatim}
\usepackage{amsthm}
\usepackage{bbm}

\makeatletter
\def\th@plain{%
	\thm@notefont{}
	\itshape 
}
\def\th@definition{%
	\thm@notefont{}
	\normalfont 
}
\makeatother

\makeatletter
\def\endthebibliography{%
	\def\@noitemerr{\@latex@warning{Empty `thebibliography' environment}}%
	\endlist
}
\makeatother

\newcommand{\rom}[1]{\uppercase\expandafter{\romannumeral #1\relax}}
\DeclarePairedDelimiterX{\norm}[1]{\lVert}{\rVert}{#1}
\DeclarePairedDelimiterX{\bnorm}[1]{\biggl\lVert}{\biggr\rVert}{#1}
\DeclarePairedDelimiterX{\abs}[1]{\lvert}{\rvert}{#1}

\renewcommand{\emph}[1]{{\textit{#1}}}

\newcommand{\ti}[1]{\tilde{#1}}

\newtheorem{definition}{Definition}
\theoremstyle{definition}

\newtheorem{example}{Example}

\def\R{\mathbb{R}}

\def\v{{\varepsilon}}

\def\F{\textsc{F}}
\def\E{\mathbb{E}}

\def\X{\mathcal{X}}
\def\Y{\mathcal{Y}}

\def\F{\mathcal{F}}

\def\M{\mathcal{M}} 
\def\I{\mathcal{I}}

\def\H{\mathcal{H}_m}

\def\A{\mathcal{A}}

\def\S{\mathcal{S}}
\def\v{\varepsilon}

\def\x{\tilde{x}}

\def\Q{\mathcal{Q}}

\def\S{\mathcal{S}}
\def\SS{\mathcal{S}'}
\def\H{\mathcal{H}}
\def\X{\mathcal{X}}
\def\Y{\mathcal{Y}}
\def\E{\mathbb{E}}

\floatname{algorithm}{Procedure}

\usepackage{xcolor}
\def\BibTeX{{\rm B\kern-.05em{\sc i\kern-.025em b}\kern-.08em
    T\kern-.1667em\lower.7ex\hbox{E}\kern-.125emX}}

\begin{document}

\title{Imitation Privacy 
%
}

\makeatletter
\newcommand{\linebreakand}{%
  \end{@IEEEauthorhalign}
  \hfill\mbox{}\par
  \mbox{}\hfill\begin{@IEEEauthorhalign}
}
\makeatother

\author{\IEEEauthorblockN{Xun Xian}
\IEEEauthorblockA{\textit{School of Statistics}\\
\textit{University of Minnesota}\\
Minneapolis, USA \\
xian0044@umn.edu}
\and
\IEEEauthorblockN{Xinran Wang}
\IEEEauthorblockA{\textit{School of Statistics} \\
\textit{University of Minnesota}\\
Minneapolis, USA \\
wang8740@umn.edu}
\and
\IEEEauthorblockN{Mingyi Hong}
\IEEEauthorblockA{\textit{Electrical and Computer Engineering} \\
\textit{University of Minnesota}\\
Minneapolis, USA \\
mhong@umn.edu}
\linebreakand 
\IEEEauthorblockN{Jie Ding}
\IEEEauthorblockA{\textit{School of Statistics} \\
\textit{University of Minnesota}\\
Minneapolis, USA \\
dingj@umn.edu}
\and
\IEEEauthorblockN{Reza Ghanadan}
\IEEEauthorblockA{\textit{Google LLC} \\
Washington, D.C., USA \\
rezaghanadan@google.com}
}

\maketitle

\begin{abstract}
In recent years, there have been many cloud-based machine learning services, where well-trained models are provided to users on a pay-per-query scheme through a prediction API.
The emergence of these services motivates this work, where we will develop a general notion of model privacy named imitation privacy. We show the broad applicability of imitation privacy in classical query-response MLaaS scenarios and new multi-organizational learning scenarios. We also exemplify the fundamental difference between imitation privacy and the usual data-level privacy. 


\end{abstract}

\begin{IEEEkeywords}
Model privacy, Machine learning services, Imitation.
\end{IEEEkeywords}

\section{Introduction}

The recent decades have witnessed substantial improvement in privacy-preserving machine learning. From cryptographic based methods, namely Homomorphic Encryption (HE)~\cite{rivest1978data} and Secure Multi-party 
Computation (SMC)~\cite{yao1986generate, goldreich1998secure}, to statistical methods, such as Differential Privacy (DP)~\cite{dwork2004privacy,dwork2006our,dwork2011differential,chaudhuri2011differentially}, local (differential) privacy~\cite{warner1965randomized,duchi2013local} and its many variations \cite{xiong2016randomized, dwork2016concentrated, mironov2017renyi,du2012privacy,wang2016relation}, most existing frameworks focus on protecting privacy at \textit{data} level. 
In data-private machine learning, a general goal is to ensure that raw data information cannot be accurately identified from learning results. 
Meanwhile, an emerging number of application scenarios involve a privately learned model, where the general public can only access its functionality through specific portals or interfaces.
Correspondingly, a new concern in contemporary machine learning is to protect a learned model instead of data. 

The `privacy' of well-trained models is significant  
in that valuable models tend to contain sensitive information extracted from the underlying raw data or domain-specific intelligence of the model specification.
For example, the leaked information from querying spam or fraud detectors~\cite{lowd2005adversarial} may be adversarially used to attack those detectors~\cite{ateniese2015hacking,fredrikson2014privacy,fredrikson2015model}. 
Also, if an adversary can mimic or even duplicate an already learned model with few queries,  service providers who trained the model may suffer from the economical loss.
A related example is algorithmic trading, where traders may easily access data, but sophisticated and successful algorithms are what truly matter. 

Model confidentiality is a particular concern in the emerging cloud-based large-scale learning services, e.g., Machine-Learning-as-a-Service (MLaaS)~\cite{alabbadi2011mobile, ribeiro2015mlaas} and privacy-aware multi-organizational learning~\cite{xian2020assisted}. These cloud-based services can either build a private model based on users' input, 
or provide prediction results using previously learned models through an Application Programming Interface (API). Often, users can only access machine learning models in a black-box fashion without knowing the model structure. 
Such pay-per-query service makes the most advanced techniques accessible to everyone, but it also put the valuable model itself as well as the training data at risk.

Recently, several work showed that an adversary could extract the servers' well-trained models through a prediction API~\cite{tramer2016stealing, chandrasekaran2018model, sethi2018data,chandrasekaran2018exploring,shi2017steal,shi2018active}. In these work, an adversary, who may or may not know the model architecture and the distributions of training data, aims to efficiently and effectively reconstruct a model $\hat{f}$ close (typically measured by out-sample prediction performances) to the API's true functionality $f$ based on a sequence of query-response pairs. Note that the typical goal of the adversary is to reconstruct a single fixed model for a specific task. From the view of approximation theory, the process of extracting a function is equivalent to a classical function approximation task. It is well known that under some mild conditions, with enough data (query-response pairs in this case), a function can be well approximated. The approximation error of the function may be viewed as a measure of privacy.

Furthermore, because the ML service providers in practice will perform more than one task using the same underlying model architectures and training data, an adversary may be more interested in imitating the \textit{modeling capability}.
In other words, an adversary aims to simultaneously imitate the performance of ML Servers on multiple tasks, with the same amount of side information from query-response pairs or limited training data. 
Such a situation would be frequently encountered in privacy-aware multi-organizations learning, e.g., Assisted Learning~\cite{xian2020assisted}, where each organization intends to assist others' learning tasks unilaterally.
Note that these tasks may include those already learned by ML Servers or those that have yet to be learned.




Our main contributions in this work are summarized below.
\begin{itemize}
    \item 
 We develop a notion of imitation privacy and its related concepts to describe the emerging model privacy issues. We discuss how imitation privacy may be used to describe various existing methods in stealing a learned model for a single task, as well as recent efforts in imitating modeling capability for multiple tasks.
    \item We discuss how imitation privacy can be used to describe the privacy leakage in privacy-aware multi-organizational learning, which does not fall into the classical query-response MLaaS scenarios. We provide several examples to illustrate how the imitation privacy may be evaluated and quantified. Particularly, we show that imitation privacy is fundamentally different from the existing data privacy, and one does not imply the other. 
\end{itemize}

The paper is organized as follows. In Section~\ref{sec_back}, we introduce some background concepts and the imitation privacy. In Section~\ref{sec_mlaas}, we explain related work in MLaaS using the imitation privacy framework. In Section~\ref{sec_al}, we elaborate on the imitation privacy in the context of multi-organizational learning, exemplified with different perspectives.

\section{Background and Formulation} \label{sec_back}
 
\subsection{Definition and Formulation}

We first introduce some concepts related to supervised learning scenarios. 
Throughout the paper, we let  $X \in \X^{n \times p} $ denote a general data matrix which consists of $n$ items and $p$ features, and $y \in \mathcal{Y}^{n}$ be a vector of labels.
The label may represent a regression response, class label, calibrated probability, or other statistics that depend on the learning tasks. For example, in a regression task, we have $\mathcal{X},\mathcal{Y} \subseteq \mathbb{R}$.
Let $x_{i}$ denotes the $i$th row of $X$.
Let $f(X)$ denote an $\R^{n}$ vector whose $i$th element is $f(x_i)$. 

\begin{definition}[Algorithm]\label{def_algo}
	A learning algorithm $\A$ is a mapping from a given dateset $X \in \R^{n \times p}$ and label vector $y \in \R^n$ to a prediction function  $f_{\A,X,y}: \R^p \rightarrow \R$.
\end{definition}

An algorithm may represent {linear regression}, {ensemble method}, {neural networks}, or a set of models from which a suitable one is chosen using model selection techniques~\cite{DingOverview}.
For example, when the least squares method is used to learn the supervised relation between $X$ and $y$, then $f_{\A,X,y}$ is a linear operator: $$\x \mapsto \x^T (X^T X)^{-1}X^T y $$ for a predictor $\x \in \mathbb{R}^p$.

\begin{definition}[Module]\label{def_module}
	A module $\M = (\A, X,y)$ is a triplet of algorithm $\A$, observed dateset $X$ and label $y$.
	For a given label vector $y \in \R^n$,  
	a module naturally induces a prediction function $f_{\A,X,y}$. 
	We simply write $f_{\A,X,y}$ as $f_{\M,y}$ to demonstrate such dependence whenever there is no ambiguity. 
\end{definition}

A module is an abstraction of a service provider in practice. 
The label may be intrinsically owned by the module or outside delivered.
In the multi-task scenario, a task-specific label ($y$) induces a corresponding prediction function.
In many classical MLaaS scenarios, both $X$ and $y$ are provided by users. In some other scenarios, e.g., the assisted learning to be elaborated in Section~\ref{sec_al}, $X$ is privately held by the module, while $y$ is provided and varies with tasks. 


\begin{figure}[H]

	\begin{center}
		\includegraphics[width=0.8\columnwidth]{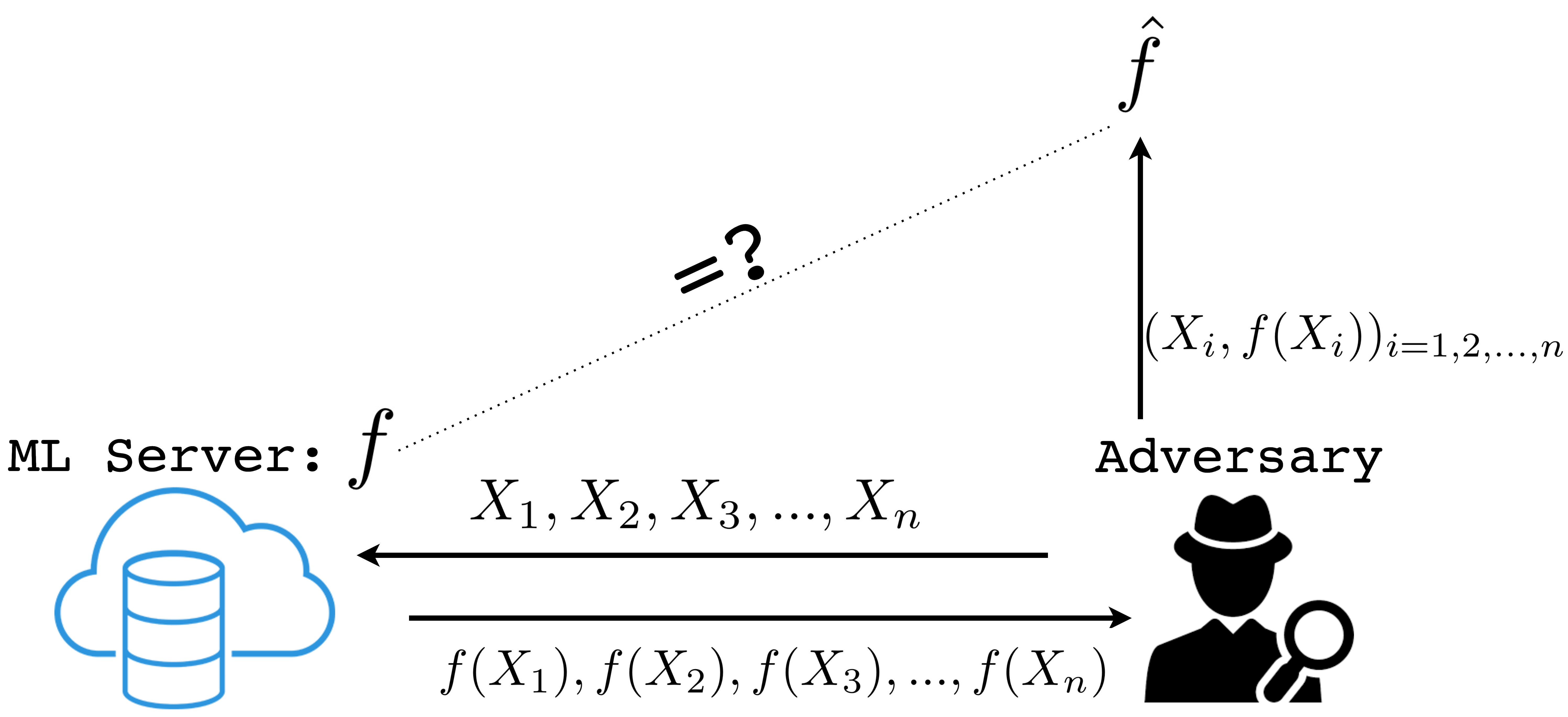}
	\end{center}
	\caption{A specific scenario of imitation: an adversary imitates a learned functionality $f$ (for a single task) through an ML Server's API.}\label{process}
\end{figure}

Now we revisit the problem introduced before by adopting our notations. 
We consider the general machine learning markets where there are multiple participants (modules), and each one of them can provide services and contribute to a specific task. 
An adversary (Alice) is denoted by $\I =(\A_0, X_0, y_0)$, where in extreme cases, $\A_0$ can be random guess strategy, $X_0$ is purely random numbers, and $y_0$ is empty. 
The ML Server's (Bob, denoted by $\M=(\A, X,y)$) model is $f_{\M,y}$ i.e., the learned model from label $y$.
The adversary aims to construct a $\hat{f}$ to mimic the functionality of the oracle model, and the closeness/accuracy is typically measured by out-sample prediction performance.

Common ML Servers offer black-box access, meaning that upon receiving queries $(X_i)_{i=1,2,...,n}$ they would only return $f(X_i)_{i=1,2,...,n}$.
Therefore, to imitate the oracle function $f$, the adversary can build a local imitating model $\hat{f}$ based on $n$ pairs of query-response. If the adversary knows any other side-information, e.g., a family containing the oracle model and the distribution of training data, constructing a imitating model would be more efficient and effective. We summarize the relevant information in the following notion of the information set. 

	\noindent\textbf{Information Set}
		An information set $\Q$ 
		 consists of triplet $ \{\ti{X}, \ti{y}, \mathcal{S} \}$, where  
		
		1) $\ti{X}$ is the data sent to the ML Server.
		
		2) $\ti{y}$ is the returned information corresponding to $\ti{X}$.
		
		3) $\mathcal{S}$ is the side information.




\noindent\textbf{Remark.} The input $\ti{X}$ can be either real-world data or random samples, either complete data entry or a partial set of features. The returned value $\ti{y}$ can be in multiple forms, such as class label, predictive probability, and confidence score, depending on different services. 
Notice that such information is different from the side information described below, as the service naturally returns it.

The $\S$ contains any possible side information for the adversary to attack that is \textit{not} obtained from interacting with the ML Server. 
For instance, the information on the ML Server's learning algorithms is one type of crucial side information. Suppose the model is a feedforward neural network. In that case, the side information could contain facts such as the number of hidden layers, the number of nodes in each layer, the type of activation/loss function, and the optimization algorithm.

Besides, the knowledge regarding the machine learning service's defensive strategy is also essential side information. In the \textit{Model Extraction Attack} literature~\cite{tramer2016stealing, chandrasekaran2018model, sethi2018data,chandrasekaran2018exploring,shi2017steal,shi2018active}, the information regarding the data used to train the service provider's model is also a kind of side information.
For example, suppose the adversary knows API's classifier is trained on Animals Image. In that case, the adversary could use the same type of images to train its own model, instead of using purely random samples.

Based on the information set $\Q$, the adversary will apply a learning algorithm $\H$ to produce an imitating model $\hat{f}$. Note that the returned values $\tilde{y}$ (e.g., labels or predicted values) in $\Q$ is induced from ML Server's underlying label $y$. We summarize the above process in the following definition.

\begin{definition}[Imitation System]
	An imitation $\I$ is a pair of query set $\Q$ and hacking algorithm $\H$ that maps $\Q$ and any label vector $y$  to a prediction function $f_{\I,y}: \R^p \rightarrow \R$ (which has the similar functionality as $f_{\M,y}$).
\end{definition}

\textbf{Remark.} Any imitation $\I$ is naturally based on an information set $\Q$ and a hacking algorithm $\H$. Whenever mentioning an imitation system, we implicitly refer to this dependence.


\subsection{Imitation Privacy}

We formally introduce the notion of imitation privacy. From an adversary's perspective, this value measures her capability of imitating. From ML Server's point of view, it measures the privacy level of his models.

	\begin{definition}[Imitation privacy]\label{def_privacy}
		The {imitation privacy} for an imitation $\I$ and module $M$ is 
		\begin{align}
		\rho_{\I, \M} = \E_{x, y \sim p_{X} \cdot p_{Y \mid X}}  \E_{\tilde{x} \sim p_{\tilde{X}}} L(f_{\I,y}(\tilde{x}), f_{\M,y}(\tilde{x}))
		\label{eq10}
		\end{align}
		where $p_{Y \mid X}$ is the task-specific distribution of label $y$,  $p_X$ is the marginal distribution of ML Server/API's training data $x$, $p_{\tilde{X}}$ is the distribution of future data $\tilde{x}$, and $L(\cdot,\cdot)$ is a loss function.
	\end{definition}
    \textbf{Remark.} 
    The loss function may depend on specific tasks.
	For example, the loss function can be scaled $L_2$ distance, i.e., $L(f(x),g(x)) = |f(x)-g(x)|^2/|f(x)|^2$ in regression, or indicator function $L(f(x),g(x)) = \mathbbm{1}(f(x) \neq g(x))$ in classification. 

	\textbf{Interpretations}. 
	The inner part of \autoref{eq10}, i.e., $\E_{\tilde{x} \sim p_{\tilde{X}}} L(f_{\I,y}(\tilde{x}), f_{\M,y}(\tilde{x}))$, describes the adversary's ability to imitate for only one specific task. The outer expectation part (expectation over joint distribution) averages such capability possibly over a set of tasks.  
	For a given $\M$, smaller $\rho_{\I,\M}$ means a closer imitation and less privacy. The minimal value $\rho_{\I,\M}=0$ is achieved at $f_{\I,y}= f_{\M,y}$ for every $y \in \Y^n$, meaning that Alice performs as well as Bob and there is `0' privacy for Bob. This can be clearly achieved when, for example, Alice holds both data and algorithm of Bob. The privacy value is typically greater than zero when Alice only holds side information such as a part of $X$ (Bob's data), a transformation of $X$, or some other data that we will demonstrate in the sequel.
	On the other hand, the value of $\rho_{\I,\M}$ is typically no larger than 1 for the above example loss function, since a trivial imitation $f_{\I,y}(x)=0$ for all $x$ leads to $\rho_{\I,\M}=1$.
	As a result, it is expected that 
	$\rho_{\I,\M}\in [0,1]$
	and it is crucial to keep a large $\rho_{\I,\M}$ for the benefit of Bob.
	
	In the definition of $\rho_{\I,\M}$, the closeness of $f_{\I,y}$ and $f_{\M,y}$ is evaluated on unobserved data (through $\E$).
	To enable easier computation, the privacy may be approximated by the training data $X=\{x_1,\ldots,x_n\}$ if $x_i$'s are assumed to be i.i.d. generated. 
	In other words, $\E$ may be replaced with $\E_n$, 
	where $\E_n f(X)=n^{-1}\sum_{i=1}^n f(x_i)$ for any measurable function $f$.
	
	The notion of imitation privacy may be extended in the following way. For two constants $\v, \delta \in [0,1]$, the module $\M$ is said to be $(\v, \delta)$-private with respect to $\I$ if with probability at most $\delta$, 
	$ 
	\rho_{\I,\M} \leq \v  \nonumber .
	$ 
	The module $\M$ is said to be $(\v, \delta)$-private with respect to a class of imitations $\mathfrak{I}$, if  
	$ 
	\inf_{\I \in \mathfrak{I}} 
	\rho_{\I,\M} \leq \v  
	$ with probability at most $\delta$. 
	The probability is due to possible randomizations of $\I$ or $\M$.

	\section{Imitation Privacy in MLaaS}\label{sec_mlaas}
    There have been several works for imitating a \textit{single} learned model in MLaas scenarios, from different perspectives. We briefly review them with the notion of imitation privacy.  The general goal of an adversary is to construct an imitating model that performs close to a target ML Server/API's true functionality, as shown in \autoref{process}.  
    The general ideology of this line of work is that more side information (available to the adversary) would lead to a more efficient and effective way to construct an imitating model. In this scenario, imitation privacy equals the average loss between ML Server's target model and adversary's imitating model evaluated on out-sample testing data (for a single task).

 \textbf{Extracting models with rich side information}.
	The concept of \textit{Model Extract Attack} was proposed in~\cite{tramer2016stealing}, where the authors developed two practical methods to precisely reconstruct a target API. 
	The first method, Equation-Solving Attacks, is suitable for extracting Logistic Regression and Multilayer Perceptrons, where the model structures are known. 
	For example, suppose that the target API is an already-trained binary logistic regression with weight $w^* \in \R^{p+1}$. For any input $\hat{x} \in \R^p$, the API will return the predicted probability, i.e., $f(\hat{x}) = \textit{logistic}(\sum_{i=1}^{p} w^*_{i} \cdot \hat{x}_{i} + w^*_0) . $
    Then, in order to recovery $w^*$, the adversary only needs to send $p+1$ random vectors and obtain their corresponding returned values and then solve $w^*$ from a linear system. 
	If the target is a decision tree, a possible solution is the Path-Finding Attack, which works in the following way. 
	For any data point $\hat{x}$ sent to the API, the response is the node's leaf-identifier that this particular input entry reaches. This identifier can trace how the data traverse in the tree and, therefore, infer the tree's structure. 
     
    Extensive experiments were conducted, and the strategies mentioned above delivered near-zero imitation privacy, i.e., the API's models were near-perfectly extracted, in most cases.
    The main reason for such excellent performance is the rich side information (knowledge of API's model class, the predicted probability, and leaf identifier) available to the adversary.


	\textbf{Extracting models with less side information}. The above two strategies have been empirically verified to deliver both efficient and straightforward solutions to extract certain types of models. However, they rely on the rich side information returned by the API, e.g., the predicted probabilities and leaf-identifiers. For classification tasks, many service providers only output predicted class labels. 
	Is it still possible to extract the model with class labels only? 
	The answer is affirmative. An adversary learning-based solution was developed in~\cite{lowd2005adversarial}, where the context is to extract a linear classifier using only the predicted class labels. The main idea is to first approximate the classifier's decision boundary, i.e., to find $\hat{x}$ such that $w^* \cdot \hat{x} + c \approx 0$, and then recovery $w^* ,c$ based on those boundary points.
	It is conceivable that the number of data (queries) needed for extracting a target model, in this case, will be much larger than the previous one. 
	The adversary typically needs to train an imitating model based on the data pairs efficiently. 
    
    \textbf{Training an imitating model}. How to effectively and efficiently train the imitating model? One crucial aspect is to select `good' training samples since those samples can accelerate the training process.
	
	 Several retraining methods named adaptive training were discussed in \cite{tramer2016stealing}.  
	 A solution is that for a given querying budget $n$, the adversary first randomly selects $m$ points and trains an imitating model. In each round of the rest $\frac{n-m}{m}$ rounds, the adversary will select points (via line search) that are close to the decision boundary, i.e. point $\hat{x}$ satisfying $w^* \cdot \hat{x} + c = 0$. The reason for choosing those points is that they are the \textit{least confident} points for the imitating model, and therefore it is essential to improve on those points. After selection, the adversary will send these data to the API and obtain the output values. Next, the adversary would update the imitating model based on the new data pairs.
 
    Efficient selection of `good' training samples is conceptually related to active learning~\cite{cohn1996active,settles2009active}. The relation was explicitly drawn in~\cite{chandrasekaran2018exploring}, where the Model Extraction Attack was cast into a Query Synthesis framework of active learning. 
	For example, to imitate a kernel SVMs model, an \textit{active selection} strategy was proposed in  \cite{chandrasekaran2018exploring} to 
	find points near the decision boundary. 
    To imitate tree-based methods, e.g., decision trees and random forests, \cite{chandrasekaran2018exploring} used the technique of importance weighted active learning (IWAL)~\cite{beygelzimer2009importance}. In particular, at $i$th round, we decide whether to query a data $x_i$ according to a Bernoulli random variable with probability $p_i \in [0,1]$ that depends on previous unlabeled examples $x_{1:i-1}$, queried labels, Bernoulli realizations, and $x_i$.
    Compared to the experimental results in~\cite{tramer2016stealing}, to achieve the same level of imitation privacy, more querying samples will be needed.
    In the above cases, the side information is only the knowledge of the target model's structure.

\textbf{Extracting models in deep learning}. 
Several deep learning techniques were adopted to imitate a target model. In fact, if the ML Server's true functionality is the deep neural network and the adversary has much side information regarding the ML Server, e.g., model structure and test/train data. Then the model extraction can be viewed as one particular case of Knowledge Distillation~\cite{hinton2015distilling}. 
    Knowledge Distillation and related work \cite{lopez2015unifying,furlanello2018born, buciluǎ2006model,zhang2018deep}  utilizes information from a complex model (teacher network) to train a smaller one (student network) with comparable accuracy. The student network will be trained on `soft target,' which is the teacher network's output.

However, work in \textit{Model Extraction} commonly assume that there is little information on the API's target model. For example, in \cite{shi2017steal}, the target model is an SVM or naive Bayes Classifier already trained from MINIST~\cite{lecun-mnisthandwrittendigit-2010}. They build a deep neural network by training on the same dataset (but with a different portion of samples) as the imitating model without knowing the API's model type, structure, and classifier parameters. They empirically showed that this would easily give near-zero imitation privacy for the API's model. 
The result is conceivable since deep neural networks have a generalized learning capabilities. 
The side information here only consists of samples from the original training data distribution, but not the knowledge of API's model class.

A technique named `knockoff nets' was introduced to extract a CNN image classifier~\cite{orekondy2019knockoff}.
	The extraction process is based on the reinforcement learning technique. In particular, each image used for querying is first assigned with a label via unsupervised clustering. Then those labeled images will be represented in a tree. For each time, the adversary will draw a node at each level according to the probability from a Softmax distribution over the nodes at the same level. After reaching the terminal node, an image will be sampled from this particular type of image pool. The adversary then sends the selected image to the API, receives the response label, and calculate the loss. 
	For the imitating model's architecture, they choose deep models such as VGG~\cite{simonyan2014very} and ResNets~\cite{he2016deep}. 
In this case, the side information is the knowledge that the target classifier is CNN trained on image data, but not any other detailed model architecture information.



\section{Imitation Privacy in Assisted Learning}\label{sec_al}

\begin{figure*}

	\begin{center}
		\includegraphics[width=1.6\columnwidth]{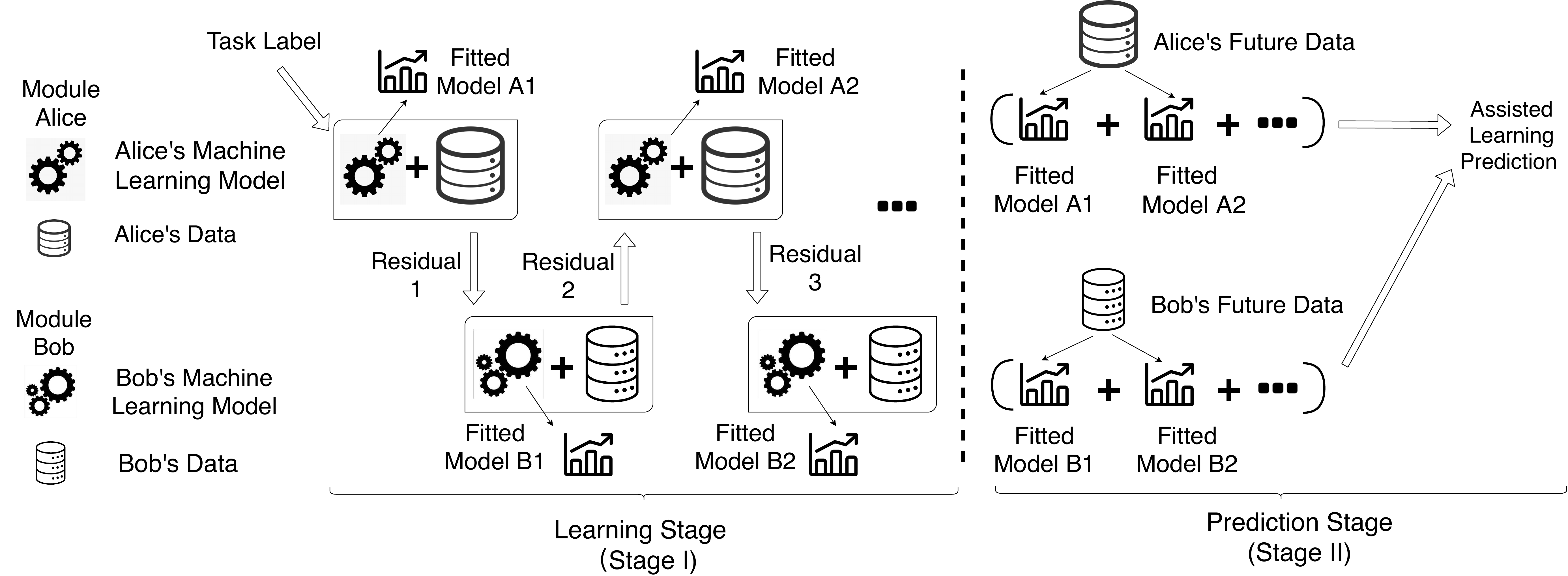}
	\end{center}
	\caption{
Illustration of the two-stage procedures for Bob to assist Alice.}\label{al_ima}
\end{figure*}

The context of Multi-Organizational Assisted Learning~\cite{xian2020assisted} concerns a group of agents with potentially heterogeneous characteristics, data, and {supervised} models. We consider the scenario where an agent (Bob) assists another agent (Alice) with her \textit{multi-task} learning, without sharing his raw data or the model that he uses. In this setting, we define the {oracle performance} as the optimal predictive loss that can be achieved from all the data aggregated from Alice and Bob  without computational constraints. 
It may be notable (or may even be counter-intuitive) that we will design a mechanism such that Alice can still achieve near-oracle performance without knowing Bob's model and raw data. 

We illustrate the idea by a simple case where the data of Alice and Bob are fully collated according to a certain data ID, and they both use regression models~\cite{xian2020assisted}.
Mathematically, Alice and Bob observe data 
features $X_A$ and $X_B$ and would like to learn regression functions $f_A$ and $f_B$ so that
$$e_{B \rightarrow A} = \E (y - f_A(X_A) - f_B(X_B))^2$$ is minimized, for a given task label $y$. Then, when a future observation  arrives, say $x_A^*$ (for Alice) and $x_B^*$ (for Bob), Bob will evaluate $f_B(x_B^*)$ and send it to Alice who will produce $f_A(x_A^*)+f_B(x_B^*)$ for prediction.
If Alice had full knowledge of  $x_B$ of Bob, she would form an augmented data  $x_{A \cup B} = [x_A, x_B]$ 
and directly learn from it to obtain an oracle error $e_{A \cup B}$.
Can Alice achieve the oracle performance  (i.e. $e_{B \rightarrow A} \approx e_{A \cup B} $) without the knowledge of $x_B$?
An affirmative solution was provided in~\cite{xian2020assisted}, which is outlined below.

As depicted in \cref{al_ima}, in \textit{State I} (learning stage), Alice first fits the data $Y$ and sends the fitted residual $e_{A,1}$ 
to Bob, who will then treat $e_{A,1}$ as his observed label and sends the corresponding fitted residuals $e_{B,1}$ back to Alice;
Alice then initiates the second round of interaction by fitting 
$e_{B,1}$ and sends a fitted residual $e_{A,2}$ to Bob, who in an analogous manner sends a new fitted residual $e_{B,2}$ back to Alice. This process is iterated $k$ times
until a stopping criterion is met. 
In \textit{State II} (prediction stage), Alice sums up all her (private) regression models  calculated in $k$ stages and forms $f_A$, and analogously so does Bob. This algorithm can be proved to approach the oracle performance~\cite{xian2020assisted}.

\begin{figure*}

	\begin{center}
		\includegraphics[height = 3.6cm, width=0.4\textwidth]{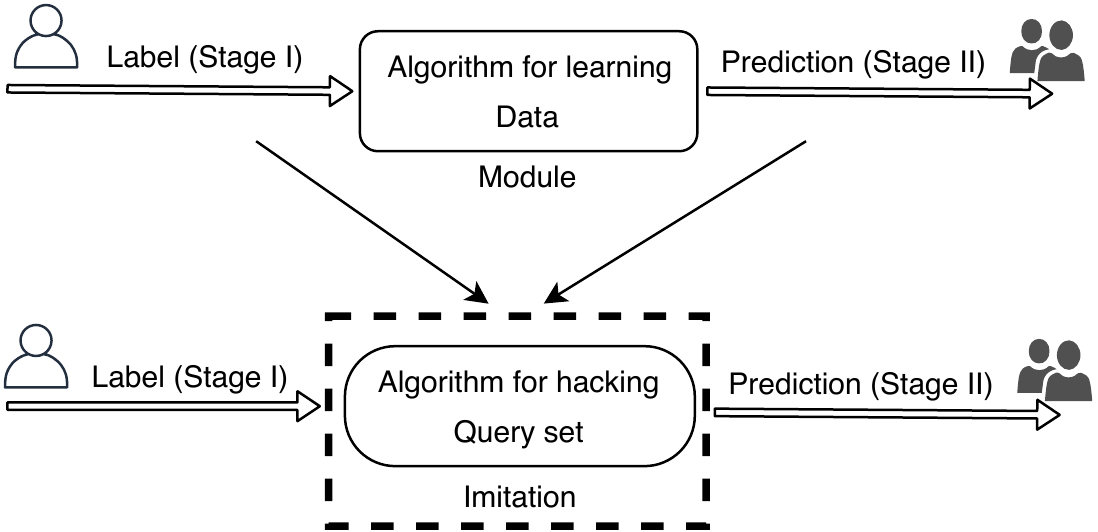}
	\end{center}
	\caption{
	A specific scenario of imitation: an Assisted Learning system (top) and its Imitation system (bottom)}\label{fig_concept}
\end{figure*}

	
The significant difference between Model Extraction in Section~\ref{sec_mlaas} and Assisted Learning is that in model extraction, the goal is to extract specific models trained on the given label. While in assisted learning,  we aim to acquire the ability of mimicking the functionality of module, which shall hold for arbitrary label.

	In the context of an assisted learning system, how do we interpret and measure the privacy for the service module Bob?
	Suppose that a user module Alice has no prior information before contacting Bob in the Assisted Learning. The only way to gain information from Bob (or ``hack the system'') is through queries at either Stage I or State II. 
	Such information is quantified below. 
	%

	There are two components of a query set. The first component is concerned with the queries at Stage II that aim to hack $f_{\M,y}(\cdot)$ \textit{for a particular $y$}. The second component is to query Stage I in order to hack the internal functionality of $\M$ itself. 
	The positive integer $k$ in our context is interpreted as the communication complexity between modules. 
	We will show by examples that a joint query to both Stage I and II is necessary to successfully imitate Bob. 
	
		\begin{example}[Algorithm leakage]
		Consider a scenario where Bob's algorithm is not available to Alice, but Bob's full data $X_B$ and the fitted response $f_{\M,y}(X_B)$ are available to Alice. 
		Suppose that there is a small fraction of data that is mismatched or overly noisy, and Bob uses a robust learning algorithm to circumvent those outliers in $X_B$ and learn an accurate model. Suppose that Alice holds a rudimentary algorithm that is sensitive to outliers. As a consequence of observing $f_{\M,y}(X_B)$, Alice would be able to identify outliers as those with  significant gaps between $f_{\M,y}(X_B)$ and $y$. 
		In this case, Bob's learning capability (in handling outliers) 
		is implicitly leaked even if Bob's algorithm is not transmitted.  
	\end{example}
	
	\begin{example}[Data leakage]
		Consider a scenario where Bob's data $X_B$ is not available to Alice, but his learning algorithm and fitted response $f_{\M,y}(X_B)$ are available to Alice. A learning algorithm will demonstrate unique information regarding the dataset, e.g. column space revealed by {linear regression} and data structure implicitly shown from {decision Tree}. In some cases, 
		Alice will be able to reverse-engineer some key statistics of Bob's data or even precise values of data, e.g. in Example \ref{lr_privacy} and \ref{dt_privacy} of the next section.

	\end{example}
	
	If Alice is not a benign user, the above query set enables her to possess useful knowledge from the service module Bob. Simply speaking, Alice aims to provide assistance to other users whatever Bob could provide, as if Alice had the algorithm and data that Bob privately holds. 
	This is formulated by the following system in parallel to a regular assisted learning system.

	\begin{definition}[Imitation system]
		An imitation system $\SS$ consists of an imitation $\I$ from an assisted learning system $\S$,  
		a learning protocol, 
		a prediction protocol, 
		and the learning procedure based on $\I$ instead of $\M$. 
	\end{definition}
An illustration of the above concepts are included in Fig.~\ref{fig_concept}. 
We now give some examples to concretely demonstrate the idea of Imitation Privacy.


	
	
	\begin{example}[Linear regression imitation privacy]
		\label{lr_privacy}
		Suppose that in an imitation, the number of Stage I queries is $k_1$,  and the numbers of Stage II queries are $ n_{1}=\cdots=n_{k}= k_2$. 
		Suppose that Bob employs a linear regression model and Alice knows about it. By  querying $k_1 \geq \min\{p, n-p\}$ random label vectors $y_{\ell}$ ($\ell=1,\ldots,k_1$) from Stage I,  Alice can obtain the column space $\textrm{span}(X_B)$ of Bob's data $X_B$ with probability one.  
		%
		Additionally, if Alice also knows about the true covariance matrix of Bob's features, 
		Alice is able to develop an imitation system with
		$\rho_{\I,\M} = o(n^{-1})$, where $n$ is the data size (or the number of rows in $X_B$).
		Note that Bob's data are never transmitted.
		
	    \begin{proof}[\textbf{Explanation}]
		    	   Suppose that in an imitation, the number of Stage I queries is $k_1$. 
		    	   Suppose that Bob employs a linear regression model and Alice knows that Bob is using a linear regression. By querying $k_1 \geq \min\{p, n-p\}$ random label vectors $y_{\ell}$ ($\ell=1,\ldots,k_1$) from Stage I,  Alice can obtain the column space $\textrm{span}(X_B)$ of Bob's data $X_B$ with probability one.  
		    	   This is because in each fitting process, Bob will project the random query $y_i$ onto $span(X_B)$, i.e.  $y_i \mapsto P_{X_B} y_i $. Hence, with $k_1=p$ fitted values in the form of $P_{X_B} y_i$, $i=1,2,...k_1$, Alice is able to uniquely identify the column space of $X_B$ with probability 1. On the other hand, with $n-p$ queries, Alice will identify the orthogonal space of $span(X_B)$, which further implies $span(X_B)$. 

		    	   However, Alice cannot obtain the Bob's data $X_B$ exactly without further side-information. 
		    	   One such side information is  the true covariance matrix of Bob.
		    	   Alice is able to develop an imitation system with
		    	   $\rho_{\I,\M} = o_p(1)$ as Bob's data size $n \rightarrow \infty$.
		    	   To see this, suppose without loss of generality that the underlying covariance of $X_B$ is an identity matrix. Let Alice arbitrarily pick up a matrix $\tilde{X} \in \R^{n \times p}$ whose column space is $span(X_B)$. 
		    	   We only need to find such a $Q$  that $ \tilde{X} = X_BQ$. 
		    	   For each label $Y_t, t=1,2,...,p$ sent in Stage I, Alice can calculate the empirical covariance between $Y$ and $X_B$, say $K_t \in \R^{p \times 1}$. By the law of large numbers, $$K_t \rightarrow_{p} cov(QX_B,Y_t) =Q cov(X_B,Y_t)$$ as $n \rightarrow \infty$. If each query $Y_t$ is in the form: $Y_t = X_B\beta_t + \eta_t$, with some fixed $\beta_t \in \R^{p \times 1}$, then Alice could solve $Q$ by letting
		    	   $$ 
		    	   \left[\begin{array}{llll} {{K}_{1}} & {{K}_{2}} & \ldots &{{K}_{p}} \end{array}\right] = Q 
		    	   \left[\begin{array}{llll} {{\beta}_{1}} & {{\beta}_{2}} & \ldots &{{\beta}_{p}} \end{array}\right] .
		    	   $$ 
		    	   As long as $\boldsymbol\beta$ is linearly independent, Alice obtains a unique $\hat{Q}_n$ that converges in probability to the true $Q$. Consequently, Alice would use $\tilde{X} \hat{Q}_n^{-1}$ as if it were $X_B$ to provide assistance, with an $o_p(1)$ imitation privacy. 
		    	   
	    \end{proof}
		%
	\end{example}
	

	\begin{example}[Decision tree imitation privacy]\label{dt_privacy}
		Suppose that Bob uses a decision tree with width $2$ and depth at least $p$, with $p$ being the number of Bob's features. Then with $k_1 \geq n $ and $k_2=\infty$, there exists an imitation such that $\rho_{\I,\M} = 0$ with high probability.
	\end{example}
	 \begin{proof}[\textbf{Explanation}]
	     Suppose that Bob uses a decision tree with width $2$ and depth at least $p$, with $p$ being the number of Bob's features. Then with $k_1 \geq n $ and $k_2=\infty$, there exists an imitation such that $\rho_{\I,\M} = o_p(1)$ as $n \rightarrow \infty$.
	    In fact, in the $i$th Stage I query, Alice sends label $y_i$ such that its $i$th entry is sufficiently large and all other entries are zero ($i=1,\ldots,n$) to identify the structure of Bob's data.
	    From the infinite Stage II queries corresponding to the $i$th Stage I query, Alice is able to reconstruct the tree built by Bob, which puts a mass at the finest neighborhood of $x_i$. Finally, Alice is able to reconstruct Bob's data up to a precision that goes to $0$ as $n$ grows. Therefore, with $k_1 \geq n$ and $k_2 = \infty$, by using the strategy described above, Alice can create an imitation system such that $\rho_{\mathcal{I},\mathcal{M}}=o_p(1)$.

	    In fact, there exist multiple ways to obtain the data structure of Bob. Next, we demonstrate another way that does not even need Stage II queries. Consider a one-dimensional case where Bob's data is $X_B^{T} = [7,1,10,5,18,9] $, and a {decision tree} with width 2 is employed. We use $x_{j}$ to denote the $j$th entry in $X_B^{T}$. Alice can obtain the structure of Bob's data by sending queries $e_i$, for $i=1,2,...,6$ to Bob in Stage I only, where $e_{i}$ is the standard basis for $\R^{6}$, and observing the fitted values $o_i$ for $i=1,2,...,6$. 
	    
	   \begin{table}[H]
	   \centering
	    	\begin{tabular}{lll}
	    		Input             & $\xrightarrow{\M}$ & Fitted Value               \\
	    		$e_1$ = {[}1,0,0,0,0,0{]} &  & $o_1$ = {[}$\frac{1}{3}$,$\frac{1}{3}$,0,$\frac{1}{3}$,0,0{]} \\
	    		$e_2$ = {[}0,1,0,0,0,0{]} &  & $o_2$ ={[}0,0,0,0,0,0{]}       \\
	    		$e_3$ = {[}0,0,1,0,0,0{]} &  & $o_3$ ={[}0,0,$\frac{1}{2}$,0,$\frac{1}{2}$,0{]}   \\
	    		$e_4$ = {[}0,0,0,1,0,0{]} &  & $o_4$ ={[}0,$\frac{1}{2}$,0,$\frac{1}{2}$,0,0{]}   \\
	    		$e_5$ = {[}0,0,0,0,1,0{]} &  & $o_5$ ={[}0,0,0,0,0,0{]}       \\
	    		$e_6$ = {[}0,0,0,0,0,1{]} &  & $o_6$ ={[}0,0,$\frac{1}{3}$,0,$\frac{1}{3}$,$\frac{1}{3}${]}
	    	\end{tabular}
	    \end{table}
	    
	    From $o_2$ and $o_5$, Alice knows that $x_{2}$, $x_{5}$ are beginning/ending points, and without loss of generality, we assume $x_{2} < x_5$. From $o_4$ and $o_1$, Alice knows $x_4$ must lie between $x_{1}$ and $x_{2}$. Similarly, the order of $x_{6},x_3,x_5$ can be inferred from $o_3$, $o_6$. Therefore, Alice has successfully recovered the structure of Bob's data, i.e., $$[x_2,x_4,x_1,x_6,x_3,x_5] = [1,5,7,9,10,18].$$ Note that without the information from Stage II, Alice can never know the exact value or even ranges of Bob's data. In fact, it is straightforward to apply such kind of strategy on complex dataset.

	 \end{proof}
	
	\begin{example}[Restricted outcome imitation privacy]
		Suppose that $y$ is generated from $$y=f(x)+\eta,$$ where $\eta\sim \mathcal{N}(0,\sigma^2)$ 
		and $f$ is randomly generated from a compact space $\mathcal{F}$ with a suitable probability measure $F$ and 
		metric $L_2(P_X)$.
		The distribution of $y$ conditional on Bob's data $X$, $p_{Y\mid X}$, 
		is Gaussian with mean $\int_{\mathcal{F}} f(X) d F$ and variance $\sigma^2$. 
		Then there exists an imitation $\I$ with 
		$$k_1=\exp\{H_{d,\v}(\mathcal{F})\}, \quad k_2=\infty,$$ 
		such that $\rho_{\I,\M} \leq \v$  with high probability. Here, $H_{d,\v}(\mathcal{F})$ denotes the Kolmogorov $\v$-entropy of $\mathcal{F}$, namely the logarithm of the smallest number of $\v$-covers of $\mathcal{F}$.
	\end{example}
	\begin{proof}[\textbf{Explanation}]
        Alice can develop the following imitation system.
		Let $f_1,\ldots,f_{k_1}$ with $k_1=\exp\{H_{d,\v}(\mathcal{F})\}$ denote the $\v$-quantizations of the function space $\F$. Suppose that $k_1$ queries are 
		constructed in such a way that  $y_j$ is generated from $f_j$, namely $y_j = f_j(x) + \eta_j$, $j=1,\ldots,k_1$.
		For any future query sent to Alice, say $y_*=f_*(x)+\eta_*$, Alice can search from the dictionary of $y_j$, $j=1,\ldots,k_1$, and find the $j$ that  
		minimizes $\norm{y_j - y_*}$. 
		Since $$n^{-1}\norm{y_j - y_*}^2 = 2\sigma^2 + \norm{f_j - f_*}_{L_2(P_X)} + o_p(1)$$ as $n\rightarrow \infty$ (assuming independent noises),  Alice would obtain such $j$ that $f_j$ is $\v$-away from $f_*$. Consequently, when Alice uses the Stage II queries corresponding to $y_j$ to assist others, she obtains an imitation privacy of $\rho_{\I,\M} \leq \v$  with high probability (for large $n$).
	\end{proof}
	

	\textbf{Relationship with Differential Privacy} One closely related concept is differential privacy.
	The main goal of differential privacy is to secure the privacy of data. 
	In the context of imitation privacy, the focus is the to secure the privacy of both the data and learning model. 
	

	A competitor Alice may attempt to imitate the learning service provided by Bob, through consecutively querying or other side information. Such an imitation, if successful, will cause an undesirable leakage of Bob's black-box learning capability.
	Below we give two examples to demonstrate that differential privacy and imitation privacy do not imply each other.

	\begin{example}[Ensured differential privacy and breached imitation privacy]
		Suppose that Bob's data $X_B$ contains $n$ i.i.d. observations of a random variable supported on $[-b,b]$. Bob can apply Laplacian mechanism to his data $X_B$ to get $\alpha$-locally deferentially private data $\tilde{X}_B$, and then releases $\tilde{X}_B$ to Alice.
		
		
		However, the above mechanism typically does not admit a non-vanishing  imitation privacy (Definition~\ref{def_privacy}, for any label distribution $p_Y$). 
		For example, suppose that Bob uses a linear regression model, then Alice can create the following imitation system with a vanishing imitation privacy. 
		For any queried $y$, Alice calculates 
		$$
		\hat{\beta}_a = (\tilde{X}_{B}^T \tilde{X}_{B} - \tau^2 I)^{-1} \tilde{X}_{B}^T y
		= (\tilde{X}_{B}^T \tilde{X}_{B} - \tau^2 I/n)^{-1} (\tilde{X}_{B}^T y/n),
		$$
		and uses $$f_{\M,y}: x \mapsto \hat{\beta}_a^T x$$ for prediction,
		where $\tau^2 = 8b^2/\alpha^2$ is the variance of the Laplacian noise that could be estimated from Stage II if not known to Alice. 
		By the law of large numbers, the above $\hat{\beta}_a$ converges in probability to the same limit as Bob's estimator $$\hat{\beta}_b = (X_B^T X_B)^{-1} X_{B}^T y.$$
		This implies a vanishing imitation privacy as the data size $n$ becomes large.
	\end{example}
	
	\begin{example}[Ensured imitation privacy and breached differential privacy]
		Suppose that a module Bob is equipped with a linear regression algorithm. Suppose that one predictor/feature is released to the public, then his dataset will not be differentially private at any privacy level. However, such direct release of partial data will only decrease the imitation privacy by a small amount. 
		Alice still can not estimate the functionality of Bob with arbitrary accuracy. 
	\end{example}


\section{Conclusion}

In this work, we define a new privacy notion at the model level, named Imitation privacy, to address the emerging privacy concerns in contemporary larges-scale cloud-based machine learning services.
We demonstrated applications of the proposed notion in both classical MLaaS and recent multi-organizational learning. Also, we discussed the fundamental differences between our notion and data privacy.
Interesting future work includes computational methods to evaluate imitation privacy for a variety of learning models on a case-by-case basis.

\balance
\bibliographystyle{IEEEtran}
\bibliography{J,ref,references_com,privacy,proposal,icml_AL,refs}

\end{document}